\documentclass[a4paper]{jpconf}
\usepackage{graphicx}
\begin{document}
\title{Spin-orbit induced spin-qubit control in nanowires}

\author{Christian Flindt$^{1,2}$,
Anders S S\o rensen$^1$ and Karsten Flensberg$^1$}

\address{$^1$ Niels Bohr Institute, Universitetsparken 5, DK-2100
Copenhagen, Denmark}
\address{$^2$ MIC -- Department of Micro and
Nanotechnology, NanoDTU, Technical University of Denmark, Building
345 East, DK-2800 Kongens Lyngby, Denmark}

\begin{abstract}
We elaborate on a number of issues concerning our recent proposal
for spin-qubit manipulation in nanowires using the spin-orbit
coupling. We discuss the experimental status and describe in
further detail the scheme for single-qubit rotations. We present a
derivation of the effective two-qubit coupling which can be
extended to higher orders in the Coulomb interaction. The analytic
expression for the coupling strength is shown to agree with
numerics.
\end{abstract}

\section{Introduction}
Gate-defined quantum dots containing only a few electrons have
been promoted as a possible candidate for solid state quantum
information processing \cite{Loss:1998}. Qubits are envisioned to
be encoded in the spin degree of freedom of the trapped electrons,
which are manipulated individually using local electron spin
resonance (ESR). Two-qubit gates are carried out by pulsing
electrically the exchange coupling between electrons in
neighboring tunnel-coupled quantum dots \cite{Burkard:1999}.
Experimentally, electric control of the exchange coupling between
two electrons in a double quantum dot was recently reported in
Ref.\ \cite{Petta:2005}. A review of the current status of quantum
computing with spins in solid state systems can be found in Ref.\
\cite{Coish:2006}.

We have recently proposed to use the spin-orbit (SO) coupling in
nanostructures as a general means to manipulate electron spins in
a coherent and controllable manner \cite{Flindt:2006}. More
specifically, we have shown how single-spin flips may be achieved
by combining the SO coupling with fast gate-induced displacements
of the electron(s), and how the SO coupling together with the
Coulomb interaction gives rise to an effective spin-spin coupling,
which is less sensitive to charge fluctuations compared to the
exchange coupling \cite{Hu:2006}.

Here, we elaborate on a number of issues related to our
proposal. First, we discuss a relevant experimental setup
consisting of a gate-defined double-dot in an InAs nanowire
\cite{Fasth:2005}. This type of setup is of particular interest to us
due to the strong SO coupling measured in InAs nanowires
\cite{Hansen:2005}. We discuss in further detail our scheme for
single-spin flips and present a derivation of the
two-spin interaction, which can be extended to
arbitrary order in the Coulomb interaction. Finally, we show that
the analytic expression for the two-spin interaction agrees with numerics.

\section{Quantum dots in nanowires}

\begin{figure}[h]
\begin{center}
\includegraphics[width=0.7\textwidth]{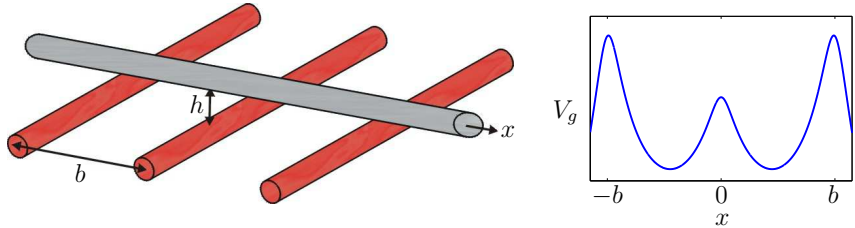}
\caption{\label{fig_setup} An InAs nanowire (light
gray) placed above three gold electrodes. The three electrodes are
used to define electrostatically a double quantum dot along the
InAs nanowire. A representative curve for the electrostatic
potential $V_g(x)$ along the wire is shown on the inset to the
right. The gold electrodes are placed at positions $x=-b,0,b$,
respectively. Typical length scales are $h\simeq 25$ nm and
$b\simeq 200$ nm. In the experiment of Ref.\ \cite{Fasth:2005} two
additional gold electrodes were used as plunger gates (not shown).}
\end{center}
\end{figure}

In the work described in Ref.\ \cite{Fasth:2005} a setup
consisting of an InAs nanowire placed above a number of gold
electrodes was successfully fabricated. The gold electrodes were
used to define electrostatically a double quantum dot within the
InAs nanowire, which was characterized using low-temperature
transport measurements, and electrostatic control of the tunnel coupling
between the two quantum dots was demonstrated. The setup is shown schematically in
Fig.\ \ref{fig_setup}.

The choice of material is interesting due to the strong SO coupling
and the large $g$-factor in InAs compared to GaAs.
In Ref.\ \cite{Hansen:2005} measurements of a
positive correction to the conductivity of InAs nanowires were
attributed to weak antilocalization arising from spin relaxation
of electrons propagating through the nanowires. This
interpretation was supported further by applying a magnetic field
that was sufficiently strong to break time-reversal symmetry,
thereby suppressing the weak antilocalization correction to the
conductivity. A spin relaxation length on the order of 200 nm was
reported, but no definitive microscopic theory for the underlying
spin-orbit coupling mechanism could be given. The exact nature of
the SO coupling in InAs nanowires is still to be fully understood
and deserves further experimental and theoretical investigation.
We expect, however, that the allowed type and strength of the SO
coupling in InAs nanowires are highly dependent on various
experimentally controllable parameters, \emph{e.g.}, the growth
direction of the nanowire.

\section{Single-spin manipulation}

We now describe how the SO coupling can be used to flip the spin
of an electron in a controllable manner. Motivated by the
structure described above we consider a one-dimensional
system\footnote{The following results are not only valid for InAs
nanowires, but more generally for gate-defined quantum dots in
one-dimensional systems with strong SO coupling.} consisting of a
single electron trapped in a gate-defined quantum dot which we
approximate with the harmonic potential
$V(x,t)=\frac{1}{2}m\omega_0^2(x-\bar{x}(t))^2$ along the
$x$-axis. We assume that the minimum position of the harmonic
potential, denoted $\bar{x}(t)$, can be varied by changing the
voltages on the gate electrodes. A static $B$-field perpendicular
to the $x$-axis splits the spin states and determines the
$z$-axis. The $x$-, $y$-, and $z$-axis define the lab frame in
which the Hamiltonian reads
\begin{equation}
H_{\mathrm{lab}}=\frac{p_x^2}{2m}+\frac{1}{2}m\omega_0^2(x-\bar{x}(t))^2+\frac{1}{2}g\mu_BB\sigma_z+\alpha
p_x\sigma_y. \label{eq_singlehamiltonian}
\end{equation}
Here $\alpha$ denotes the strength of the SO coupling. The
particular form of the SO coupling is assumed to arise from lack
of inversion symmetry in the $yz$-plane \cite{Levitov:2003}.
In order to get a feeling for the SO coupling it is useful to introduce the SO length
$l_{\mathrm{so}}\equiv\hbar/m\alpha$:
Without an applied $B$-field, a spin along the $z$-axis is flipped
after having traveled the distance $\pi\l_{\mathrm{so}}/2$ along
the $x$-axis. Another important length
scale is the oscillator length defined as
$l_0\equiv\sqrt{\hbar/m\omega_0}$.

It is convenient to work in a frame that follows the SO induced
rotation of the spin. We shall refer to this frame as the rest
frame. The Hamiltonian in the rest frame $H_{\mathrm{rest}}$ is
obtained by the unitary transformation
$H_{\mathrm{rest}}=UH_{\mathrm{lab}}U^{\dagger}$, where
$U=\exp{(i\sigma_y(x-\bar x(0))/l_{\mathrm{so}})}$, and we find
\begin{equation}
H_{\mathrm{rest}}=\frac{p_x^2}{2m}+\frac{1}{2}m\omega_0^2(x-\bar{x}(t))^2+\frac{1}{2}g\mu_BB\left[\cos{\left(\frac{2(x-\bar{x}(0))}{l_{\mathrm{so}}}\right)}\sigma_z-\sin{\left(\frac{2(x-\bar{x}(0))}{l_{\mathrm{so}}}\right)}\sigma_x\right].
\label{eq_restframe}
\end{equation}
The static $B$-field in the lab frame is rotating in the rest
frame of the spin as it travels along the $x$-axis. In the
following we work in a regime, where the equilibrium position
$\bar{x}(t)$ is slowly changed on the time-scale of the orbital
degree of freedom, while fast on the time-scale of the spin,
\emph{i.e.}, $g\mu_BB/\hbar\ll(1/l_0)d\bar{x}(t)/dt\ll\omega_0$.
This allows us to trace out the orbital degree of freedom by
projecting $H_{\mathrm{rest}}$ onto the oscillator ground state,
whereby we arrive at an effective spin Hamiltonian reading
\begin{equation}
H_{\mathrm{spin}}=\frac{1}{2}\tilde{g}\mu_BB\left[\cos{\left(\frac{2(\bar{x}(t)-\bar{x}(0))}{l_{\mathrm{so}}}\right)}\sigma_z-\sin{\left(\frac{2(\bar{x}(t)-\bar{x}(0))}{l_{\mathrm{so}}}\right)}\sigma_x\right]
\end{equation}
with the renormalized $g$-factor given by
\begin{equation}
\tilde{g}=g\left\langle e^{2i(x-\bar{x}(0))/\l_{\mathrm{so}
}}\right\rangle=ge^{-(\l_0/\l_\mathrm{so})^2}.
\label{eq_spinhamiltonian}
\end{equation}
Here the brackets denote an average with respect to the oscillator
ground state.

\begin{figure}[h]
\begin{center}
\includegraphics[width=0.25\textwidth]{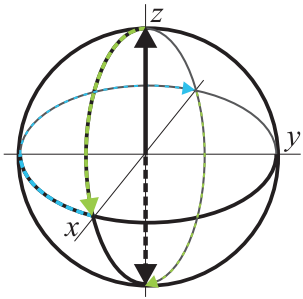}\hspace{2pc}
\begin{minipage}[b]{25pc}\caption{\label{fig_ramsey}
Bloch sphere in the lab frame with the full black
arrow denoting the initial spin state. The three colored dashed
lines indicate the rotations of the spin during the spin flip
scheme. The green lines correspond to the fast displacements of
$\bar{x}(t)$, while the blue line in the $xy$-plane corresponds to
the rotation around the static $B$-field. The dashed black arrow
denotes the final spin state.}
\end{minipage}
\end{center}
\end{figure}

Using the spin Hamiltonian given in Eq.\
(\ref{eq_spinhamiltonian}) we can manipulate the spin by changing
the equilibrium position $\bar{x}(t)$. Below we describe a scheme
for spin flips, which does not rely on any resonance conditions as
in previous studies \cite{Rashba:2003,Golovach:2006}. Instead, our
scheme relies on fast (on the time scale of the spin) and large
(on the order of $l_{\mathrm{so}}$) changes of $\bar{x}(t)$
obtained by controlling the potentials on the gate electrodes.
Considering a spin being in an eigenstate of the spin Hamiltonian
at $t=0$, \emph{i.e.},
$H_{\mathrm{spin}}=\frac{1}{2}\tilde{g}\mu_BB\sigma_z$, the scheme
reads:
\begin{enumerate}
\item Fast displacement of $\bar{x}(t)$: $\bar{x}(0)\rightarrow \bar{x}(0)+\pi
l_{\mathrm{so}}/4$. This rotates the $B$-field into the $x$-axis
of the rest frame.
\item Free evolution of the spin now governed by the spin Hamiltonian
$H_{\mathrm{spin}}=-\frac{1}{2}\tilde{g}\mu_BB\sigma_x$ for a time
span $\Delta t=\hbar\pi/\tilde{g}\mu_BB$. This rotates the spin in
the rest frame by $\pi$ around the $x$-axis.
\item Fast return of $\bar{x}(t)$: $\bar{x}(0)+\pi
l_{\mathrm{so}}/4\rightarrow \bar{x}(0)$. This returns the
$B$-field to the initial position in the rest frame, pointing
along the $z$-axis.
\end{enumerate}
After the three steps, a spin initially prepared in an eigenstate
of $H_{\mathrm{spin}}=\frac{1}{2}\tilde{g}\mu_BB\sigma_z$ has been
flipped into the other eigenstate. For realistic experimental
parameters one finds an estimated time for the spin flip process
on the order of 0.1 ns \cite{Flindt:2006}. A graphical
interpretation of the spin flip process (as seen in the lab frame)
is given in Fig.\ \ref{fig_ramsey}. We note that the spin Hamiltonian given in Eq.\ (\ref{eq_spinhamiltonian})
allows for rotations of the spin to any point on the Bloch sphere.

\section{Two-spin manipulation}

The SO coupling also couples spins in neighboring quantum dots. The form and the strength of the coupling can be
found by considering two electrons, each described by a (rest
frame) Hamiltonian $H^{(i)}_{\mathrm{rest}}$ of the form given in
Eq.\ (\ref{eq_restframe}) with different values of $\bar{x}_i$,
$i=1,2$, for each of the two quantum dots. The orbital degrees of
freedom are coupled due to the Coulomb interaction between the
electrons, which we expand using $1/|x_2-x_1|\simeq
1/d-\delta/d^2+\delta^2/d^3$, where we have introduced the
distance between the quantum dots $d\equiv \bar{x}_2-\bar{x}_1>0$
and assumed that
$d\gg\delta\equiv(x_2-\bar{x}_2)-(x_1-\bar{x}_1)$. Retaining only
the term in the expansion of the Coulomb interaction that couples
the positions of the electrons, the two-particle Hamiltonian reads
\begin{equation}
H=H^{(1)}_{\mathrm{rest}}+H^{(2)}_{\mathrm{rest}}-\frac{e^2}{2\pi\varepsilon_0\varepsilon_rd^3}(x_1-\bar{x}_1)(x_2-\bar{x}_2).
\label{eq_twoparticlehamiltonian}
\end{equation}
For the two-spin coupling temporal variations of $\bar{x}_i,$
$i=1,2,$ are not necessary.

An effective two-spin Hamiltonian $H_{\mathrm{eff}}$ can be found
using imaginary time formalism. The two-particle Hamiltonian given
in Eq.\ (\ref{eq_twoparticlehamiltonian}) is written $H=H_0+H'$,
where $H_0$ denotes the Hamiltonian of the two oscillators and $H'\equiv H-H_0$. We
define $e^{-\beta
H_{\mathrm{eff}}}\equiv\mathrm{Tr_{osc}}(e^{-\beta H})/Z_0$,
$Z_0\equiv \mathrm{Tr_{osc}}(e^{-\beta H_0})$, where
$\mathrm{Tr_{osc}}$ denotes a (partial) trace over the two
oscillators. Introducing the operator $\hat{U}(\beta)\equiv
e^{\beta H_0}e^{-\beta H}$, and the thermal average of an operator
$A$ with respect to $H_0$, $\langle A\rangle_0\equiv
\mathrm{Tr_{osc}}(Ae^{-\beta H_0})/Z_0$, we write $e^{-\beta
H_{\mathrm{eff}}}=\langle \hat{U}(\beta)\rangle_0$. Using the formal expression
$\hat{U}(\beta)=T_{\tau}\exp\left(-\int_{0}^{\beta}d\tau\hat{H}'(\tau)\right)$,
where $T_{\tau}$ denotes the (imaginary) time-ordering operator and
$\hat{H}'(\tau)\equiv e^{\tau H_0/\hbar}H'e^{-\tau H_0/\hbar}$ is
the interaction picture representation of $H'$ (in imaginary
time),
we can in principle calculate the effective two-spin Hamiltonian
$H_{\mathrm{eff}}$ to any order in $H'$. The first non-vanishing
term that couples the two spins arises from the expansion of
$\hat{U}(\beta)$ to third order in $H'$ and has the form
$\tau_{xx}\sigma_x^{(1)}\sigma_x^{(2)}$, where $\tau_{xx}$ is to
be determined. Concentrating on this term, we find
\begin{equation}
e^{-\beta H_{\mathrm{eff}}}\simeq \ldots+
\frac{(eg\mu_BB)^2}{8\pi\varepsilon_0\varepsilon_rd^3}\int_{0}^{\beta}\int_{0}^{\beta}\int_{0}^{\beta}d\tau_1d\tau_2d\tau_3\mathcal{G}(x_1-\bar{x}_1,\tau_1-\tau_2)\mathcal{G}(x_2-\bar{x}_2,\tau_2-\tau_3)\sigma_x^{(1)}\sigma_x^{(2)}+\ldots,
\end{equation}
where we have assumed that the spin degrees of freedom evolve much
slower than the orbital part, and $\hat{\sigma}_x^{(1)}$ and $
\hat{\sigma}_x^{(2)}$ are thus taken to be time-independent. The
correlation function $\mathcal{G}(x_i,\tau)$ is defined as
\begin{equation}
\mathcal{G}(x_i,\tau)\equiv\left\langle T_{\tau}
\sin\left(\frac{2\hat{x}_i(\tau)}{l_{\mathrm{so}}}
\right)x_i\right\rangle_{0},
\end{equation}
and can be evaluated using linked cluster theory. We find
\begin{equation}
\mathcal{G}(x_i,\tau)=\frac{l_0^2}{l_{\mathrm{so}}}e^{-(l_0/l_{\mathrm{so}})^2}\Big[\Theta(\tau)\Big(n_B(\beta\hbar\omega_0)e^{\omega_0\tau}+(1+n_B(\beta\hbar\omega_0))e^{-\omega_0\tau}\Big)
+(\tau\rightarrow-\tau)\Big],
\end{equation}
where $\Theta(\tau)$ is the Heaviside step function and
$n_B(x)=1/(e^{x}-1)$. Collecting all terms and carrying out the
triple integral, we find
\begin{equation}
e^{-\beta
H_{\mathrm{eff}}}=\ldots+\frac{e^2}{4\pi\varepsilon_0\varepsilon_r}\frac{2\l_0^4(g\mu_BB)^2}{\l_{\mathrm{so}}^2(\hbar\omega_0)^3d^3}e^{-2(l_{\mathrm{so}}/l_0)^2(1+n_B(\beta\hbar\omega_0
))}F(\beta\hbar\omega_0)\sigma_x^{(1)}\sigma_x^{(2)}+\ldots
\end{equation}
with $F(x)=[x(\cosh(x)-2)+\sinh(x)]/2\sinh^2(x/2)$. Having solved
the finite-temperature ($\beta<\infty$) problem, we let
$\beta\rightarrow\infty$, and identify
\begin{equation}
\tau_{xx}=-\frac{e^2}{4\pi\varepsilon_0\varepsilon_r}\frac{2\l_0^4(\tilde{g}\mu_BB)^2}{\l_{\mathrm{so}}^2(\hbar\omega_0)^2d^3}
\label{eq_tau}
\end{equation}
in agreement with our previous work \cite{Flindt:2006}. The
imaginary time formalism outlined here allows for calculation of
the coupling to higher order in the Coulomb interaction and corrections due to a finite temperature.

\section{Numerics}

\begin{figure}[h]
\begin{center}
\includegraphics[width=0.44\textwidth]{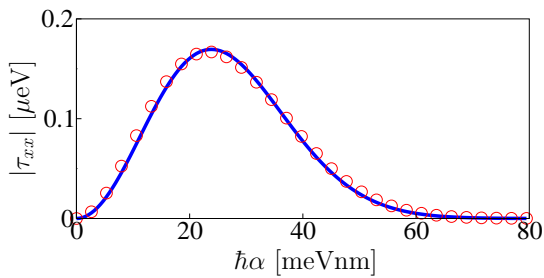}\hspace{2pc}
\begin{minipage}[b]{18.7pc}\caption{\label{fig_RashbaFit}
Two-qubit coupling $\tau_{xx}$ as a function of the SO coupling
$\alpha$. Parameters are $\varepsilon_r=15.15$, $m=0.027m_e$,
$g=14.8$, $B=160$ mT, $l_0=80$ nm ($\hbar\omega_0\simeq 0.4$ meV)
and $d\simeq 0.75$~$\mu$m. Circles indicate numerical results,
while the full line shows Eq.\ $(\ref{eq_tau})$.}
\end{minipage}
\end{center}
\end{figure}

In Ref.\ \cite{Flindt:2006} we used a numerical implementation of
the two-particle Hamiltonian in Eq.\
(\ref{eq_twoparticlehamiltonian}) to study the coupling
$\tau_{xx}$ in case of non-harmonic confining potentials. Here we
use the numerical implementation to calculate $\tau$ as a function
of the SO coupling strength $\alpha$. In Fig.\
\ref{fig_RashbaFit} we show a comparison of numerical results for
$\tau_{xx}$ using nearly harmonic confining potentials and Eq.\
(\ref{eq_tau}). The numerical results agree well with the analytic
results, and as expected the coupling depends quadratically on
$\alpha$, \emph{i.e.}, $\tau_{xx}\propto\alpha^2$ for small values
of $\alpha$ (corresponding to $l_{\mathrm{so}}\gg l_{0}$), while
it for large values of $\alpha$ is dominated by the renormalized
$g$-factor which drops off exponentially with increasing $\alpha$.

\section{Conclusion}

We have elaborated on our recent proposal for spin-qubit manipulation using the SO coupling
in nanostructures. We have discussed an experimental setup with strong SO coupling
which may be relevant for realizing our proposal. We have described in detail how single-spin rotations may be carried
out using fast displacements of the electron(s), and we have derived an expression for
the effective two-spin interaction using imaginary-time formalism. Finally, we have shown that the analytic result for the two-spin
interaction agrees with numerics.

\ack

The authors thank G Burkard, X Cartoixa, W A Coish, A Fuhrer and J
M Taylor for illuminating discussions during the preparation of
this manuscript.

\section*{References}

\end{document}